\documentclass[nofootinbib]{revtex4}
\usepackage{latexsym,comment,verbatim}
\usepackage{amssymb}
\usepackage{amsfonts}
\usepackage{amsmath}
\usepackage{color}
\usepackage{graphicx}
\usepackage{bm}

\def\ber{\begin{eqnarray}}
\def\eer{\end{eqnarray}}
\def\beq{\begin{equation}}
\def\eeq{\end{equation}}

\def\ed{\end{document}}

\begin{document}

\title{Testing general relativity by means of ringlasers}
\author{Angelo Tartaglia}
\email{angelo.tartaglia@polito.it}
\affiliation{Politecnico di Torino and ISMB, Corso Duca degli Abruzzi 24, 10129 Torino, Italy}
\author{Angela Di Virgilio}
\email{angela.divirgilio@pi.infn.it}
\affiliation{INFN Pisa, Polo Fibonacci Largo B. Pontecorvo, 3, 56127 Pisa, Italy}
\author{Nicol\'{o} Beverini}
\email{nicolo.beverini@unipi.it}
\affiliation{Physics Department, University of Pisa and INFN Pisa, Polo Fibonacci Largo B. Pontecorvo, 3, 56127 Pisa, Italy}
%
\author{ Jacopo Belfi}
\email{jacopo.belfi@pi.infn.it}
\affiliation{INFN Pisa, Polo Fibonacci Largo B. Pontecorvo, 3, 56127 Pisa, Italy}
\author{Matteo Luca Ruggiero}
\email{matteo.ruggiero@polito.it}
\affiliation{Politecnico di Torino and INFN, Corso Duca degli Abruzzi 24, 10129 Torino, Italy}
%
%
\begin{abstract}
  The paper discusses the optimal configuration of one or more ring lasers to be used for measuring the general relativistic effects of the rotation of the earth, as manifested on the surface of the planet. The analysis is focused on devices having their normal vector lying in the meridian plane. The crucial role of the evaluation of the angles is evidenced. Special attention is paid to the orientation at the maximum signal, minimizing the sensitivity to the orientation uncertainty. The use of rings at different latitudes is mentioned and the problem of the non-sfericity of the earth is commented.
\end{abstract}


\maketitle

\date{\today}

\newpage

\section{Introduction}

Ring Lasers (RL) are top sensitivity devices able to measure absolute
rotations. The principle of operation of a ring laser is based on the Sagnac effect \cite{sagnac}. RLs are very reliable instruments, with large bandwidth and very high duty cycle. The most advanced RLs are indeed used for accurate metrology in geophysics (rotational seismology),
and in geodesy for monitoring the fast variations of the Earth rotation rate.

For a laboratory on Earth, the signal of a RL is proportional to the instantaneous\footnote{Actually the measured rotation rate is over a time interval, corresponding to the lifetime of a photon in the cavity of the ring. The latter is however in the order of $10^{-3}$ s, much smaller than the typical times of other time depending phenomena in the lab and on the planet altogether.}
norm of the vector sum of the diurnal rotation rate of the planet, $\vec{%
\Omega}_\oplus$, and the local rotation rate of the device, $\vec{\Omega}_l$%
; the two components together may be called \textit{kinematic rotations}. In principle the superposition of local and global kinematic rotations produces a time varying signal, since local rotations are referred to a non-inertial reference frame, so that the rotation rate with respect to distant inertial observers turns out to cyclically depend on time. In many practical applications the latter time dependence is \textit{de facto} negligible.

When the effects of non-Newtonian gravity are included, an additional
contribution may appear; let us call it $\vec{\Omega}_{gr}$. If General
Relativity (GR) is used, $\vec{\Omega}_{gr}$ is in turn the sum of two
contributions: the Lense-Thirring drag term $\vec{\Omega}_{LT}$ and the de
Sitter geodetic precession $\vec{\Omega}_{dS}$. If the RL is carried on a
vehicle the dominant term is $\vec{\Omega}_l$; in a laboratory fixed to the
ground $\vec{\Omega}_\oplus$ prevails; the GR terms are $\sim10^{-14}$
rad/s, nine orders of magnitude below the Earth rotation rate. As for $\vec{%
\Omega}_l$, in an Earth based laboratory it is either negligible or known and modelled
so that it can be accounted for and subtracted. The present best sensitivity of a RL is $\sim10^{-13}$ rad/s
in one day of integration time \cite{wettzell}, not far from the threshold
to be crossed in order to detect the GR terms.

In short, the response of the
RL is a beat frequency $f$ proportional to the scalar product between the
total angular rotation vector and the area vector, $A\hat{n}$, of the ring: $%
f = S(\vec{\Omega}_\oplus+\vec{\Omega}_l+\vec{\Omega}_{gr})\cdot \hat{n}$.
The proportionality factor $S$ is called scale factor and depends on the
geometry of the ring. It is $S = \frac{4 A}{\lambda P}$, where $A$ is the
area and $P$ the perimeter of the ring, $\lambda$ is the wavelength of the
light of the Laser. With an appropriate construction and location of the
apparatus and for long enough integration time we may assume $\langle
\Omega_l \rangle$ to be negligible, even with respect to the GR terms, and
other effects to be modelled and subtracted accurately, so that, in the
framework of General Relativity, we write $f = S(\vec{\Omega}_\oplus+\vec{%
\Omega}_{LT}+\vec{\Omega}_{dS})\cdot \hat{n}$.\newline
The purpose of the GINGER experiment (Gyroscopes IN GEneral Relativity) is
to measure the GR components of the gravitational field of the Earth at $1\%$
or better accuracy level, by means of an array of ring-lasers. In 2011
a first proposal was presented based on an octahedral configuration \cite{PR2011}%
. The three-dimensional array would permit to reconstruct the modulus of the
total angular rotation vector in the laboratory. The GR terms in this scheme
would be evaluated by subtracting the Earth rotation rate measured
independently by the International Earth Rotation and Reference Systems Service (IERS), $\vec{\Omega}_{IERS}$. The
proposed approach would require long term stability and very high accuracy,
since it would be necessary to subtract the contribution of $\vec{\Omega}%
_\oplus$, which, as said, is about nine orders of magnitude bigger than the GR
terms.

So far the gravitomagnetic field of the Earth has been measured by
spaceborne experiments, being the present accuracy limit $\sim 5\%$ \cite%
{lares}. The experimental goal to measure $\Omega_{LT}$  down to $1\%$,
remains an important challenge. GINGER would provide the
first measurement of the General Relativistic features of the gravitational
field, on the surface of the Earth (not considering the gravitational
redshift). Though not in free fall condition, it would be a direct local
measurement independent from the global distribution of the gravitational
field, which is the principle difference with the space experiments where the result is the consequence of an averaging of the effects along whole orbits.\newline

In the following we shall discuss the ways an actual measurement based on ring lasers can be done, evidencing criticalities and the role of physical and geometrical parameters and the related uncertainties.

\section{Ring lasers for retrieving a general rotation vector}
\label{II}

As we have already written in the Introduction, the beat frequency $f$ of a RL is proportional to the flux of a total rotation vector $\vec{\Omega}_t$ across the area of the ring. In general we may write:

\beq
f=S\vec{\Omega}_t \cdot \hat{n}
\label{flusso}
\eeq
where $\hat{n}$ is the unit vector perpendicular to the plane of the ring (provided, of course, that it is contained in a plane).
 If we wish to fully recover $\vec{\Omega}_t$ from the measurement of frequencies we need in principle three independent rings, which form a local three-dimensional reference frame, as it was proposed in \cite{PR2011}. If we have reasons to think that a couple of rings may be oriented so that the plane of their $\hat{n}$'s contains $\vec{\Omega}_t$, the problem becomes bi-dimensional and two rings are enough; we shall comment on this later, but let us assume for the moment that this is the case.

\begin{figure}[htbp]
\begin{center}
\includegraphics[scale = 0.3]{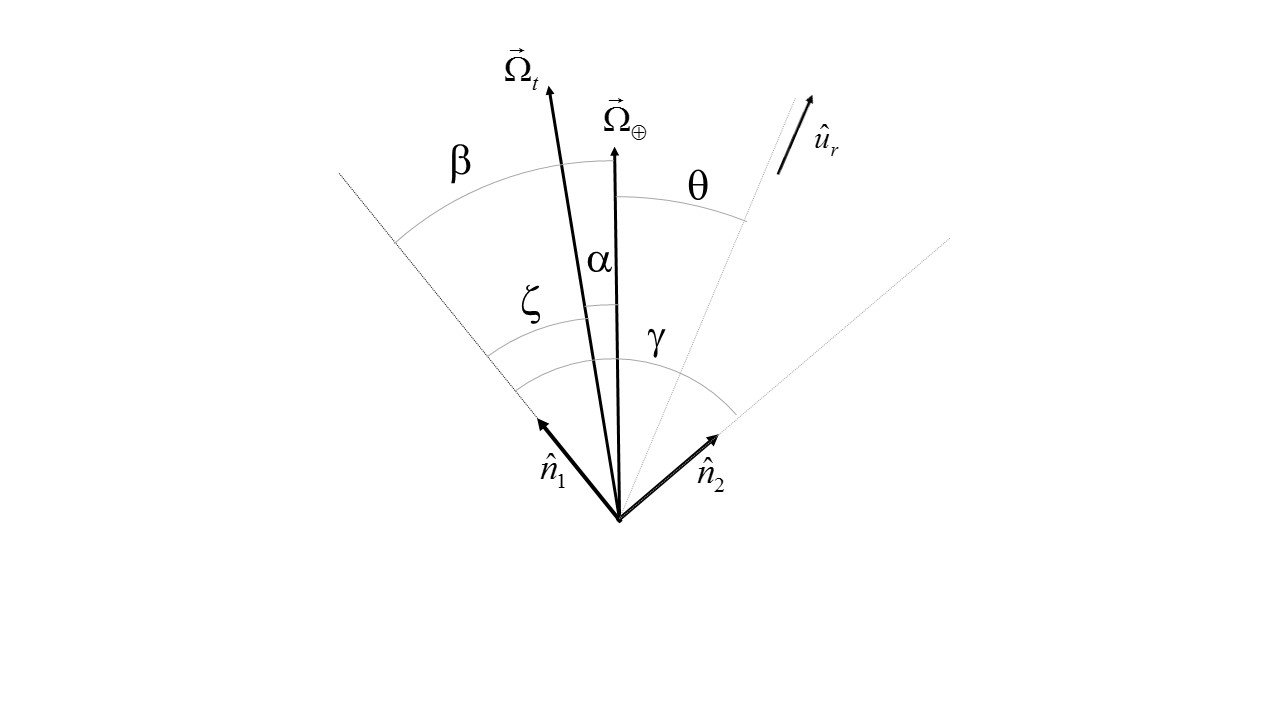}
\end{center}
\caption{A schematic, pictorial, not on scale, view of the orientation of the
kinematic $\vec{\Omega}_\oplus$ and the total effective rotation vector $%
\vec{\Omega}_{t}$ in the meridian plane. The unit area vectors of the two rings, $\hat{n%
}_1$ and $\hat{n}_2$ are also shown. The angle
between $\vec{\Omega}_t$ and $\vec{\Omega}_\oplus$ is $\protect\alpha$; $%
\protect\beta$ and $\protect\gamma$ are respectively the angles between ring
1 and $\vec{\Omega}_\oplus$, and ring 2 with respect to ring 1. Angle $\zeta$ is between $\vec{\Omega}_t$ and $\hat{n}_1$.}
\label{twoDt}
\end{figure}

 Making the scalar products explicit and calling $\gamma$ the angle between $\hat{n}_1$ and $\hat{n}_2$ (see Fig. \ref{twoDt}) we may write:

 \ber
 f_1&=&S_1 \Omega_t \cos{\zeta} \nonumber \\
 \label{system0} \\
 f_2&=&S_2 \Omega_t \cos{(\gamma-\zeta)} \nonumber
 \eer
Here $\zeta$ is the angle between $\vec{\Omega}_t$ and $\hat{n}_1$.
System (\ref{system0}) may be transformed into:

\ber
\frac{f_2}{f_1}&=&\frac{S_2}{S_1} \frac{\cos(\gamma-\zeta)}{\cos\zeta} \\
\Omega_t&=&\frac{f_1}{S_1 \cos\zeta}=\frac{f_2}{S_2 \cos(\gamma-\zeta)}
\eer

Provided $\gamma$ is known (directly measured), we may solve and obtain:

\ber
\tan\zeta&=&\frac{S_1 f_2-S_2 f_1 \cos\gamma}{S_2f_1\sin\gamma} \nonumber\\
\label{generale} \\
\Omega_t&=&\frac{\sqrt{S_1^2 f_2^2+S_2^2 f_1^2-2S_1S_2f_1f_2\cos\gamma}}{S_1 S_2 \sin\gamma} \nonumber
\eer

Of course everything simplifies if we may assume $S_1=S_2=S$ and $\gamma=\pi/2$. It would then be:

\begin{subequations}
\ber
\tan\zeta&=&\frac{f_2}{f_1} \label{tanz}  \\
S\Omega_t&=&\sqrt{f_2^2+f_1^2} \label{Omega2}
\eer
\end{subequations}

In order not to get in trouble with signs, we should specify a couple of assumptions: 1) $\vec{\Omega}_t$ is in between $\hat{n}_1$ and $\hat{n}_2$; 2) it is $0<\gamma\leq \pi/2$.

Either in the form (\ref{generale}) or (\ref{tanz}) (\ref{Omega2}) the two rings give $\vec{\Omega}_t$ without any reference to the composition of the vector and specifying the orientation in the meridian plane with respect to themselves.

\section{General Relativity: The RingLaser signal}
\label{signal}

The metric of the external space-time of a spherical rotating mass is written in the simplest form in the
reference frame of an inertial observer located at infinity and at rest with
respect to the center of the source of gravity. It is also convenient to
work in weak field approximation, where terms down to the smallest
interesting contribution are kept. The smallest term we keep is linear in
the angular momentum of the central mass $J$. The line element, using space
"polar" coordinates, is:
\begin{equation*}
ds^{2}=\left( 1-2\frac{m}{r}\right) c^{2}dt_{\ast }^{2}-\left( 1+2\frac{m}{r}%
\right) dr^{2}-r^{2}d\theta ^{2}-r^{2}\sin ^{2}\theta d\phi _{\ast }^{2}+4%
\frac{j}{r^{2}}\sin ^{2}\theta \left( cdt_{\ast }\right) \left( rd\phi
_{\ast }\right)
\end{equation*}
Variables marked by pedix $_{\ast }$ will change when passing to the final
frame. It has been assumed that $m^2/r^2<j/r^2$ and negligible. It is
\begin{eqnarray*}
m &=&G\frac{M_{\oplus }}{c^{2}}\simeq 4.43\times10^{-3}  \text m  \notag \\
j &=&G\frac{J}{c^{3}}=G\frac{I}{c^{3}}\Omega _{\oplus }\simeq 1.75\times10^{-2}  \text m^2  \notag
\label{manfj}
\end{eqnarray*}
The last assumption includes the hypothesis that the earth is a rigid body
whose relevant moment of inertia is $I$. The numerical values, when considering the surface of the earth, i.e. $r=R=6.373\times10^6$ m, confirm that the approximation adopted is correct.

The measurement is intended to be performed in a terrestrial laboratory, so
it is appropriate to rewrite the line element in its reference frame. This
is made through two steps \cite{JMPD}:

\begin{itemize}
\item rotation of the axes at the angular velocity of the earth $%
\vec{\Omega}_{\oplus}$;

\item boost at the peripheral speed of the earth whose absolute value is $%
V=\Omega _{\oplus }R\sin \theta $ where $\theta $ is the colatitude of the
laboratory and $R$ is the (average) radius of the earth.
\end{itemize}

In the process, we keep an approximation level consistent with the weak field
hypothesis, extended to kinematical rotation terms. The size of the latter is expressed by the ratio $\Omega_\oplus R/c\sim 1.55\times 10^{-6}$.

The result is:
\begin{eqnarray}
ds^{2} &=&\left( 1-2\frac{m}{r}\right) c^{2}dt^{2}-\left( 1+2\frac{m}{r}%
\right) dr^{2}-r^{2}d\theta ^{2}  \notag \\
&&-\left( 1+2\frac{r^{2}\Omega _{\oplus }^{2}}{c^{2}}\sin ^{2}\theta \right)
r^{2}\sin ^{2}\theta d\phi ^{2}  \label{ellinea} \\
&&+2\left( 2\frac{j}{r^{2}}-r\frac{\Omega _{\oplus }}{c}-2m\frac{\Omega
_{\oplus }}{c}\right) \sin ^{2}\theta \left( cdt\right) \left( rd\phi \right)
\notag
\end{eqnarray}

For short we write
\begin{equation*}
g_{0\phi }=\left( 2\frac{j}{r^{2}}-r\frac{\Omega _{\oplus }}{c}-2m\frac{%
\Omega _{\oplus }}{c}\right) \sin \theta
\end{equation*}
The frame is non-inertial and comoving with the laboratory; the origin
remains in the center of the Earth. Considering a null line-element (i.e. a
light ray: $ds=0$) from (\ref{ellinea}) we deduce the coordinated travel
time of flight element $dt$. Assuming a path closed in the laboratory (it is
not closed for an inertial external observer) and integrating along the path
once to the right ($d\phi >0$), once to the left ($d\phi <0$), then
subtracting the two results, we arrive to the difference in the coordinated
times of flight (expressed in arbitrary coordinates):

\begin{equation*}
\delta t=-2\oint \frac{g_{0i}}{g_{00}}dx^{i}
\end{equation*}

It is possible to convert the result to the proper time of the observer at
rest in the lab, $\tau$, just multiplying by $\sqrt{g_{00}}$ at his/her
position

\begin{equation*}
\delta \tau =-2\sqrt{g_{00}}\oint \frac{g_{0i}}{g_{00}}dx^{i}.
\end{equation*}

Considering the symmetry of the problem we may interpret $g_{0\phi }$ as the
only non-zero component of a three-vector $\vec{h}$ aligned with the axis of
rotation of the Earth so that (introducing the unit vector $\hat{u}_{l}$
aligned with the trajectory of light) the formula becomes:
\begin{equation*}
\delta \tau =\frac{2}{c}\sqrt{g_{00}}\left\vert \oint \frac{\vec{h}\cdot
\hat{u}_{l}}{g_{00}}dl\right\vert
\end{equation*}
The quantity is an observable, i.e. a true scalar: the same for any observer.

In a ring laser the time of flight asymmetry is converted into a difference
in the frequency of stationary light beams and one obtains a beat frequency:
\begin{equation*}
f=\frac{2c}{\lambda P}\sqrt{g_{00}}\left\vert \oint \frac{\vec{h}\cdot \hat{u%
}_{l}}{g_{00}}dl\right\vert
\end{equation*}
The line integral may be transformed into a flux using Gauss's theorem
(classically it would be called Stoke's theorem). If the change of the
values of the curl of $\vec{h}/g_{00}$ across the area of the closed
integration path is negligible, the result becomes simply
\begin{equation*}
f=\frac{2cA}{\lambda P}\vec{\nabla}\wedge \left( \frac{\vec{h}}{\sqrt{g_{00}}%
}\right) \cdot \hat{u}_{n}
\end{equation*}%
where $A$ is the area contoured by the beams, $P$ is the length of the path,
$\lambda $ is the wavelength in the active cavity and the curl is evaluated
in any point within the ring. Besides the physical and geometrical
parameters, to be controlled experimentally, the signal depends on three
quantities: $m$ (proportional to the mass of the source $M_{\oplus }$), $I$
( contained in $G\frac{\mathit{j}}{c^{3}}$), and $\Omega _{\oplus }$. After
a few manipulations, the expected signal becomes

\begin{subequations}
\begin{eqnarray}
f &=& \frac{4A}{\lambda P}\left[ \vec{\Omega}_{\oplus }-2\frac{m}{r} \Omega
_{\oplus }\sin \theta \hat{u}_{\theta }+G\frac{I\Omega _{\oplus }}{
c^{2}r^{3}}\left( 2\cos \theta \hat{u}_{r}+\sin \theta \hat{u}_{\theta
}\right) \right] \cdot \hat{u}_{n}  \label{base} \\
&=& S\left( \vec{\Omega}_{\oplus }+\vec{\Omega}_{dS}+\vec{\Omega}%
_{LT}\right) \cdot \hat{u}_{n}  \label{baseb}
\end{eqnarray}
\end{subequations}

As we see, according to GR, $\vec{\Omega}_t$ is the sum of three vectors, all contained in one plane (the meridian plane), thus explaining the special attention paid to the two rings system in the previous section.

\begin{figure}[htbp]
\begin{center}
\includegraphics[scale=0.3]{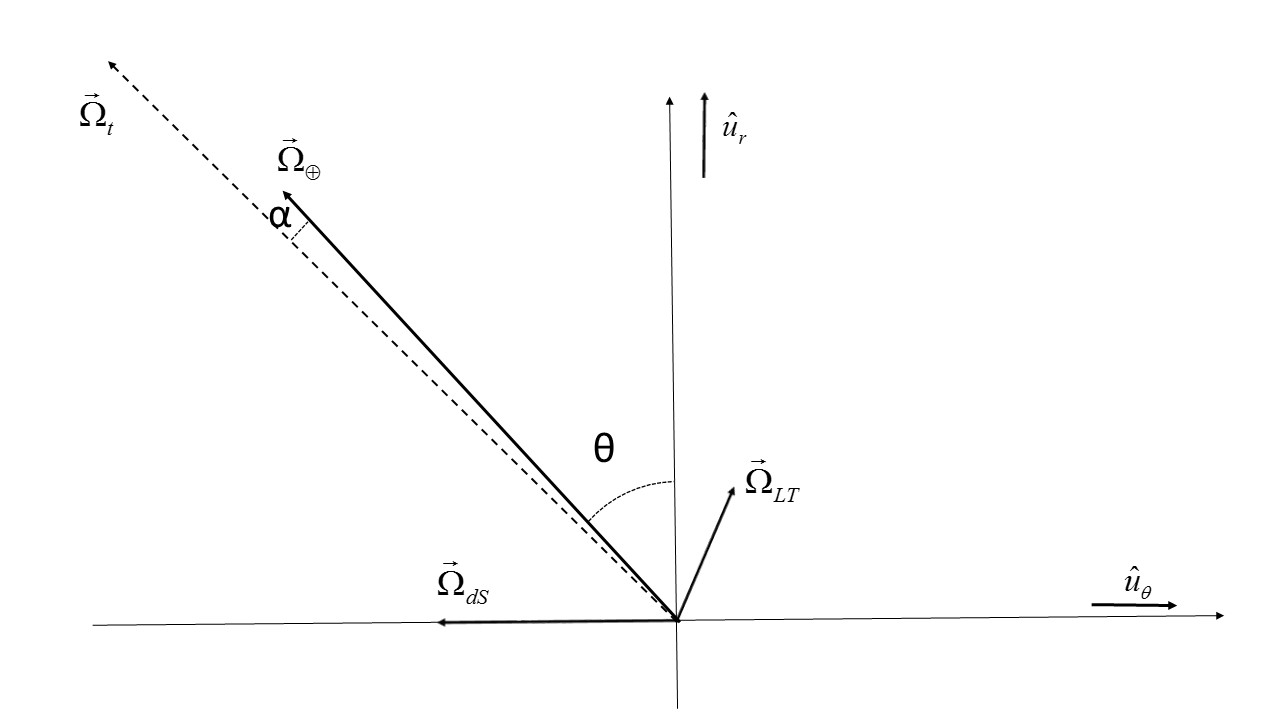}
\end{center}
\caption{The mutual orientation of the angular velocity of the Earth $\vec{\Omega}_\oplus$ and of the two GR effective rotations $\vec{\Omega}_{LT}$
and $\vec{\Omega}_{dS}$ is represented. The sum $\vec{\Omega}_t$ of all rotation vectors is also shown. The amplitude of $\vec{\Omega}_\oplus$ is down-scaled by about 10 orders of magnitude. The graph is not on scale.}
\label{Orientation}
\end{figure}

Let us introduce the angle $\beta$ between the direction of the axis of the
Earth and the axis of the ring (see Fig. \ref{Orientation}). Then the relevant dot products become:

\begin{eqnarray*}
\hat{u}_{\Omega_{\oplus}}\cdot \hat{u}_{n} &=&\cos \left( \beta \right) \\
\hat{u}_{r}\cdot \hat{u}_{n} &=&\cos \left( \beta -\theta \right) \\
\hat{u}_{\theta }\cdot \hat{u}_{n} &=&\sin \left( \beta-\theta \right)
\end{eqnarray*}

Introducing the shorthand notations

\begin{subequations}
\begin{eqnarray}
a &=& 2\frac{m}{R}  \label{amr} \\
b &=& \frac{GI}{c^2 R^3}  \label{bir}
\end{eqnarray}
\end{subequations}

Eq. (\ref{base}) is converted into:

\begin{equation}
f=S\Omega _{\oplus }\left\vert \cos \left( \beta \right) -\left( a-b\right)
\sin \theta \sin \left( \beta-\theta\right) +2b\cos \theta \cos \left( \beta
-\theta \right) \right\vert  \label{frequenza}
\end{equation}

The absolute value bars $\|$ have been introduced just to remember that the
frequency is of course always a positive quantity. The factor in front of
the bar is the scale factor $S$ of the RL. It is also important to remark
that $\Omega_\oplus$ acts as a global multiplication factor. Eq.s (\ref{base}) and (\ref{frequenza}), unlike Eq.s (\ref{system0}), present the expected orientation of the RL referring to the external frame formed by $\hat{u}_r$ and $\hat{u}_\theta$; this is manifested by the presence of the colatitude $\theta$ and the angle $\beta$ (in the configuration represented in Fig. \ref{Orientation} the two angles have opposite signs).

As stated at the beginning, we have treated throughout the earth as a sphere, but we know that our planet is not a sphere. Without entering into the details of the geoid, we should better treat the earth as an ellipsoid: how would this better approximation affect formulae like (\ref{base}) and (\ref{frequenza})? We can recall that the shape parameter of the terrestrial reference ellipsoid is $\simeq 0.003$. It should produce locally a deviation of the vertical direction (i.e. the local gravitoelectric field) from the radial direction (which appears in the formulae) in the order of $\sim 10^{-3}$ rad at most. Such a deviation affects the GR terms in Eq. (\ref{base}) on the corresponding amount of $1$ part in $10^3$ at most, so being below the target accuracy for the experiment.

\section{Confronting the experiment with the theory}
\label{expTheor}

Going back to Eq. (\ref{base}) we may express $\vec{\Omega}_{\oplus}$ in
terms of $\hat{u}_r$ and $\hat{u}_{\theta}$. We then obtain an explicit formula for the modulus of $\vec{\Omega}_t$:

\beq
\Omega_t = \Omega_{\oplus}\sqrt{(1+(a-b)(a-b+2)\sin^2\theta +
4b(b+1)\cos^2\theta)}  \label{OmT}
\eeq

Equating (\ref{OmT}) to the second equation in (\ref{generale}) we get a relation between the experimental quantities and a combination of the parameters of the theory:

\beq
\Omega_{\oplus} \sqrt{1+(a-b)(a-b+2)\sin^2\theta +
4b(b+1)\cos^2\theta}=\frac{\sqrt{S_1^2 f_2^2+S_2^2 f_1^2-2S_1S_2f_1f_2\cos\gamma}}{S_1 S_2 \sin\gamma}
\eeq

Remember that $a$ and $b$ are expected to be of the order of $10^{-9}$, so that we may keep the only first order corresponding terms:

\beq
 1+2b\cos^2\theta+(a-b)\sin^2\theta \simeq \frac{\sqrt{S_1^2 f_2^2+S_2^2 f_1^2-2S_1S_2f_1f_2\cos\gamma}}{\Omega_{\oplus} S_1 S_2 \sin\gamma}
\label{solparz}
\eeq

If, for simplicity, we assume that the Earth is
spherical with an internal uniform mass distribution, the relationship
between $I$ and $M_\oplus$ is: $I=\frac{2}{5}M_{\oplus}R^{2}$ ($b=a/5$). For
the real Earth the numerical factor is closer to $1/3$ rather than $2/5$ ($%
b=a/6$)(see the Appendix).

Apart from the modulus of $\vec{\Omega}_t$ the theory tells us also the orientation of the vector in the meridian plane: it will be at an angle $\alpha$ with respect to the axis of the Earth.
When the ring has its axis oriented as $\vec{\Omega}_t$, the signal reaches its maximum value:

\beq
f_{max}=S\Omega_t
\eeq

Angle $\alpha$ can be obtained applying the maximum condition to Eq. (\ref{frequenza}). In fact, starting from Eq. (\ref{frequenza}), using $\beta$ as the independent variable, calling $\alpha$ the value
of the angle at the maximum, we find:

\begin{equation}
\tan\alpha=\frac{(a-3b)\sin\theta\cos\theta}{(3b-a)\sin^2\theta -2b-1}
\label{alpha}
\end{equation}

Considering the orders of magnitude, the first order approximation is:

\begin{equation}
\alpha\simeq (3b-a)\sin\theta\cos\theta  \label{sima}
\end{equation}

Posing $\beta=\alpha$ in (\ref{frequenza}), then using (\ref{alpha}) or (\ref{sima}), we may obtain from (\ref{frequenza}) the same $\Omega_t$ as the one written in (\ref{OmT}).

In order to establish a correspondence between the predictions of the theory and the results obtained from a couple of rings, let us remark that Eq. (\ref{frequenza}) can also be written putting $\beta=\zeta+\alpha$. Looking at (\ref{generale}), let us consider a configuration where ring 1 is aligned at the maximum; this corresponds to $\zeta=0$. If the second ring is horizontal (i.e. $\hat{n}_2\parallel \hat{u}_r$), it is $\gamma=\alpha+\theta$ (see Fig. \ref{Orientation}). We may write:

\beq
\tan\alpha=\frac{\tan\gamma-\tan\theta}{1+\tan\gamma\tan\theta}
\eeq

Finally, calling in (\ref{alpha}) or (\ref{sima}), we arrive at:

\beq
\frac{(a-3b)\sin\theta\cos\theta}{(3b-a)\sin^2\theta -2b-1}=\frac{\tan\gamma-\tan\theta}{1+\tan\gamma\tan\theta}
\label{secex}
\eeq

or

\beq
a-3b\simeq\frac{\tan\theta-\tan\gamma}{(1+\tan\gamma\tan\theta)\sin\theta \cos\theta}
\label{secap}
\eeq

The end point of this process is reached putting (\ref{secap}) and (\ref{solparz}) in a system:

\ber
a-3b&\simeq&\frac{\tan\theta-\tan\gamma}{(1+\tan\gamma\tan\theta)\sin\theta \cos\theta} \nonumber \\
\label{sist} \\
1+2b\cos^2\theta+(a-b)\sin^2\theta &\simeq& \frac{\sqrt{S_1^2 f_2^2+S_2^2 f_1^2-2S_1S_2f_1f_2\cos\gamma}}{\Omega_{\oplus} S_1 S_2 \sin\gamma}  \nonumber
\eer

Choosing $a$ and $b$ as unknowns, all other parameters must be measured. In principle system (\ref{sist}) can be solved. The formal result is:

\ber
a&=&-\frac{3}{2}+\frac{\tan\theta-\tan\gamma}{1+\tan\theta\tan\gamma}\frac{2\cos^2\theta-\sin^2\theta}{2\sin\theta \cos\theta}+\frac{3}{2}\frac{\sqrt{S_1^2f_2^2+S_2^2f_1^2-2S_1S_2f_1f_2\cos\gamma}}{\Omega_{\oplus}S_1S_2\sin\gamma} \label{aaa}\\
b&=&-\frac{1}{2}-\frac{(\tan\theta-\tan\gamma)\tan\theta}{2(1+\tan\theta \tan\gamma)}+\frac{\sqrt{S_1^2f_2^2+S_2^2f_1^2-2S_1S_2f_1f_2\cos\gamma}}{2\Omega_{\oplus}S_1S_2\sin\gamma}
\label{bbb}
\eer

The practical difficulty with this solution is that it is composed of strongly differing parts, scaling over at least eleven orders of magnitude (if the aim is a $1\%$ accuracy in $a$ and $b$), so requiring a corresponding accuracy in all parameters, including $\vec{\Omega}_{\oplus}$.

Going back to system (\ref{sist}), we see that the first equation is apparently purely geometrical, without calling in $\vec{\Omega}_{\oplus}$. Now the basis is the measurement of $\gamma$, which in fact is equivalent to the direct measurement of $\alpha$; remember that $\gamma$ is the angle between $\vec{\Omega}_t$ (i.e. the direction of the maximum along which the first ring is oriented) and $\hat{u}_r$ (which is perpendicular to the plane of the second ring).
Using the first equation only, it would even be unnecessary to have the second ring, but of course the angles must be measured with an accuracy better than $1$ nrad and the result would be the combination $a-3b$. To have $a-3b$, instead of the two parameters separately, would not be a problem since we know the relation between the two. Of course the same results may be obtained from (\ref{aaa}) and (\ref{bbb}).

\section{Various configurations}

The configuration first considered in 2011, as recalled in the Introduction, has been the octahedron \cite{PR2011}. This configuration has been extensively discussed in previous papers; it
measures the three components of $\vec{\Omega}_t$ in all three spatial directions
and reconstructs the norm of the vector combining together different
measurements. This approach allows the comparison of different co-located
rings, giving the possibility of precisely measuring the systematics of the
laser. If the orientation of the octahedron with respect to the rotation axis of the Earth and the meridian plane are not given, all information that can be retrieved by the experiment is contained in the norm of $\vec{\Omega}_t$ confronted with the theory. If also the external orientation is given, the full $\vec{\Omega}_t$ vector, including the angle with respect to the axis of the Earth, is obtained.

In general, using multiple independent rings (two, three or more) has several advantages: the statistics would
be improved since the shot noise of each ring, in the set of many, is
independent from the others. Co-locating more than three rings would be a powerful tool to keep the systematics
of an experimental apparatus under control. An array of at least four
co-located rings would have the very interesting feature that the angular
rotation vector could be reconstructed with different combinations $3$ by $3$%
. The comparison of different results would give information on the
systematics of the lasers. Redundancy would be allowed, which is always welcome in this kind of experiment.

The weak point is that the detection of the GR terms requires the knowledge of $\vec{\Omega}_\oplus$ which in practice is provided by IERS and $\Omega_{IERS}$ is given with an uncertainty too high, as for now, to allow to
reconstruct the Lense-Thirring effect at the $1\%$ accuracy level. In fact, the Length of Day
(LoD) is measured with different methods by the IERS, but,
in the best case, with a $10\div15$ $\mu s$ error. This is compatible with a $%
10\%$ test, more or less; improvements are not foreseen in the next five years plan (IERS
Annual Report 2014 \cite{IERS2014}). It is however true that prolonging the measurement time would reduce the uncertainty (1 order of magnitude in 10 days), but of course one must insure the stability of the apparatus over the whole extension of the run.

\subsection{The output of a single ring and the RL at the maximum signal}

As we have already seen, a single ring measures the projection of the total $\vec{\Omega}_t$ on a
direction perpendicular to the plane containing the ring. In principle a
single ring could give all the information (first equation in (\ref{system0})) besides the orientation of the total vector with respect to the axis of the Earth, but the knowledge of the absolute angle $\zeta$ between the normal to the ring and the direction of the maximum signal is required. The normal to the ring is assumed to lie in the meridian plane. Using the first approximation of Eq. (\ref{OmT}) it is:

\beq
2b\cos^2\theta+(a-b)\sin^2\theta \simeq \frac{f}{S\Omega_{\oplus} \cos\zeta}-1
\label{uno}
\eeq

A special case is obtained when the ring is oriented to the maximum signal. It is then $\zeta=0$ and $f=f_{max}$. An advantage of this configuration is that, being in a maximum condition, the sensibility to orientation inaccuracy is second order: an uncertainty of the order of a $\mu rad$ affects the frequency at the $prad/s$ level. Furthermore the orientation with respect to the axis of the Earth is directly given by the theory, Eq. (\ref{alpha}) or (\ref{sima}).

In any case using the simple proportionality relation mentioned at the end of Sect. \ref{expTheor},
the unknown is reduced to one (for instance $a$) and we may solve for it.

\subsubsection{Horizontal ring}

Another special case is a horizontal ring (normal in the meridian plane and aligned with the local Newtonian field). The output is obtained from Eq. (\ref{frequenza}) putting $\beta=\theta$:

\beq
f_h=S\Omega_\oplus (1+2b)cos\theta
\label{oriz}
\eeq
whence the gravito-magnetic parameter $b$ immediately stems.

The advantage of this configuration is that, unlike other orientations, it corresponds to an angle materially defined in the laboratory. Unfortunately this simplification is not so strong as it looks: horizontality is not an extremal condition, so it has to be reached with an accuracy of the same order of magnitude as the one required for the GR term. In practice the maximum tolerable deviation from the horizontal plane is of the order of a $prad$. We must mention that the horizontality condition is affected by the shape of the geoid, which has not  regular surface, according to the comments we have put at the end of section \ref{signal}. Here too, however, the effect on the GR terms as such is negligible. Unfortunately the isolation of the Lense-Thirring term requires the subtraction of a contribution of the order of $S\Omega_\oplus$ which must be known in 1 part in $10^{12}$ and this is the real reason for the $prad$ requirement mentioned above.

\subsubsection{Ring containing the direction of the terrestrial axis}
It is worth remarking that Eq. (\ref{base}) tells also that GR terms could,
in principle, be obtained keeping the normal in the meridian plane, but
orienting the ring so that its plane contains the direction of the axis of
the Earth (orthogonality condition between the axis and the normal to the
plane of the ring). In that configuration the kinematic frequency
would be zero. Unfortunately, this method is not viable for two reasons: 1)
ring-lasers must be operated with a bias in order to avoid the locking of
the two counter-propagating modes (and the GR terms alone would probably be
too small to give the necessary bias); 2) the accuracy required in the
alignment in order to insure that the ring keeps its normal in the meridian
plane becomes extremely severe.  The latter statement may be verified by an
example: considering a square ring-laser $6$ m in side, and in the same time with its normal
perpendicular to the axis of the Earth, a tilt of $30$ $prad$ (from the zero
kinematical contribution orientation) would mimic the expected GR signal. In
general a single ring requires a control of its absolute orientation and,
based on the above arguments, the practically viable solution is the ring oriented
at the maximum signal. A single ring parallel to the axis of the Earth has no practical application: the RL
does not work properly, and the requirements on the accuracy of $\beta$ become
exceedingly severe. These considerations hold also in the case of a pair of nested rings perpendicular to each other ($\gamma=\pi/2$), where one is oriented to the maximum: the second ring would not work.

\subsection{Rings at different latitudes}

Keeping $a$ and $b$ as separate unknowns, the dependence on the co-latitude can be exploited. A couple of measurements performed by rings in laboratories located at different latitudes could give both unknowns.

The general equations are:

\begin{eqnarray*}
f_{\theta =\theta _{1}} &=&S_1\Omega _{\oplus }\left\vert \cos \left( \beta_1
\right) +\left( a-b\right) \sin \theta _{1}\sin \left( \theta _{1}-\beta_1
\right) +2b\cos \theta _{1}\cos \left( \beta_1 -\theta _{1}\right) \right\vert
\\
f_{\theta =\theta _{2}} &=&S_2 \Omega _{\oplus
}\left\vert \cos \left( \beta_2 \right) +\left( a-b\right) \sin \theta
_{2}\sin \left( \theta _{2}-\beta_2 \right) +2b\cos \theta _{2}\cos \left(
\beta_2 -\theta _{2}\right) \right\vert
\end{eqnarray*}

If both rings are oriented to the maximum, it is:

\begin{eqnarray}
f_{max 1} &\simeq &S_1\Omega _{\oplus }\allowbreak \left( 1+2b\cos ^{2}\theta
_{1}+\left( a-b\right) \sin ^{2}\theta _{1}\right) \nonumber \\
\label{latit} \\
f_{max 2} &\simeq &S_2\Omega _{\oplus }\allowbreak \left( 1+2b\cos ^{2}\theta _{2}+\left( a-b\right) \sin ^{2}\theta
_{2}\right) \nonumber
\end{eqnarray}

The formal solution of the system (\ref{latit}) is:

\begin{eqnarray}
a&\simeq&-\frac{3}{2}+\frac{1}{2\Omega_\oplus}\frac{f_{max2}S_1-f_{max1}S_2}{S_1S_2(\cos{2\theta_1}-\cos{2\theta_2})}\allowbreak
+\frac{3}{2}\frac{f_{max2}S_1\cos{2\theta_1}-f_{max1}S_2\cos{2\theta_2}}{\Omega_\oplus S_1S_2(\cos{2\theta_1}-\cos{2\theta_2})}  \nonumber \\
\label{solu} \\
b&\simeq&\frac{\sin^2{\theta_1}-\sin^2{\theta_2}}{\cos{2\theta_1}-\cos{2\theta_2}}\allowbreak
+\frac{f_{max1}S_2\sin^2{\theta_2}-f_{max2}S_1\sin^2{\theta_1}}{\Omega_\oplus S_1S_2(\cos{2\theta_1}-\cos{2\theta_2})} \nonumber
\end{eqnarray}

Here too, a useful combination obtainable from (\ref{solu}) is

\beq
a-3b\simeq2\frac{f_{max2}S_1-f_{max1}S_2}{\Omega_\oplus S_1S_2(\cos{2\theta_1}-\cos{2\theta_2})}
\label{a-3b}
\eeq

The advantage of this equation is that $\Omega_\oplus$ (actually its inverse) appears only as a global multiplier, which means that the requirement on its absolute accuracy is much less severe than in the cases where it has to be subtracted from something else.

Another possibility worth mentioning is with a ring to the maximum at latitude $\theta_1$ and the other horizontal at latitude $\theta_2$. The equations are:

\begin{eqnarray}
f_{max1}&\simeq& S_1\Omega_\oplus \left[1+b(2\cos^2\theta_1-\sin^2\theta_1)+a\sin^2\theta_1\right] \nonumber \\
f_{h2}&\simeq& S_2\Omega_\oplus (1+2b)\cos\theta_2 \nonumber
\label{mh}
\end{eqnarray}

The formal solution is now:

\begin{eqnarray}
a &\simeq& -\frac{3}{2} - \frac{f_{max1}}{S_1\Omega_\oplus \sin^2\theta_1} +f_{h2} \frac{\sin^2\theta_1-2\cos^2\theta_1}{2S_2\Omega_\oplus \cos\theta_2\sin^2\theta_1}  \nonumber \\
b &\simeq& -\frac{1}{2}+\frac{f_{h2}}{2S_2\Omega_\oplus \cos\theta_2} \nonumber
\label{smh}
\end{eqnarray}

Finally the convenient combination:

\beq
a-3b \simeq \frac{f_{max1}S_2 \cos\theta_2-f_{h2}S_1}{\Omega_\oplus S_1S_2 \cos\theta_2\sin^2\theta_1}
\label{solhor}
\eeq

Again the impact of the accuracy on $S\Omega_\oplus$ is reduced, but the second ring needs to be laid in the horizontal plane within a $prad$ or so.

\subsection{Two rings in the same place}

This case has already been treated in Sect.s \ref{II} and \ref{expTheor}. A constraint for this configuration is to insure that the normals to both rings lay in the meridian plane. It is a manageable condition, since the meridian is a symmetry plane; the additional contributions coming for an out-of-the plane component of $\vec{\Omega}_t$ would be proportional to the cosine of an angle $\phi$ whose value on the plane would be $0$. In practice, as for the orientation to the maximum, an uncertainty of $10^{-6}$ rad on $\phi$ would affect the measured frequencies at the $10^{-12}$ rad/s level only.

The retrievable information is contained in system (\ref{sist}) and its solutions. Recalling the discussion made in Sect. \ref{expTheor} we remark again that in principle the only ring at the maximum could be enough (Eq. (\ref{secap})), provided one is able to measure the angle to the radial direction $\gamma$, then exploiting the relation between $a$ and $b$.

\subsection{Frequencies}

In the discussions presented so far, always appear frequencies $f$, either
as expected values obtained from the knowledge of the other physical
parameters and the angles, or as input data to find the GR quantities $a$
and $b$. It must be recalled that the $f$'s are not the measured frequencies
given by the RLs system, since in that experimental output other effects are
contained too. The measured frequencies also account for the movements of
the axis of the Earth, originated from external perturbations; the
rotational motions of the crust of the planet; the local movements of the
ground and of the laboratory originating from various possible causes, etc.
Last but not least there are the instabilities of the laser, down to the
shot noise. The random components tend to average to zero, prolonging the
duration of each experimental run; the other contributions need to be
identified and then can be described as best as possible by appropriate
modeling. In any case whatever is not Earth rotation and GR must carefully
be subtracted from the raw data in order to obtain the frequencies to be
used in our formulae. The cleaning process must of course be as accurate as
the required final accuracy in the $f$'s.

\section{Conclusion}

We have analyzed and discussed various possible configurations and orientations of up to three RLs, located either in one place or at different latitudes. We have laid down the basic equations to be used in the various cases and shown the level of accuracy required, if the GR effects are aimed at. A convenient choice is to have one ring oriented to the maximum signal so that the orientation accuracy can be relaxed to approximately $1$ $\mu$rad. Apart from the above, the critical parameters are the angles, which, with the exception of the orientation at the maximum, should be known within the prad.

So far, the physical principles and constraints have been treated, laying down the fabric of possible experiments and the constraints to be abided by in order to give relevant results. Next come the measurement strategies and all the features of a real experiment. The behaviour of the laser and of the resonant loop need to be discussed, then the behaviour of the mirrors used to obtain the closed path for light, back scattering effects, etc.. The mechanical and thermal stability of the whole setup have to be taken into account, and so on. It is clear that, for practical reasons, it will be convenient to work with a redundancy of rings, allowing to mutually control and quantize the intrinsic uncertainties. All these aspects are under consideration and will be treated in a technical paper, now in preparation.

For sure, the experiment is not an easy one, but both the general considerations we have presented, and the ongoing technological trends tell us that the objective of using ring laser arrays for fundamental physics is a viable one and the GINGER project and collaboration moves on the right path. An additional bonus for this type of measurements is that they also provide a good amount of important information on the behaviour of the geophysics and geodesy of our planet.

\appendix

\section{General relativistic parameters of the planet Earth}
\label{abterra}

From the general definitions we may express $b$ as a function of $a$: $b=k%
\frac{a}{2}=k\frac{GM_{\oplus }}{c^{2}R}$. The present knowledge
about the size and shape of the Earth is thoroughly exposed in \cite{torge} and the most recent data are in \cite{NASA}. It is:

\begin{eqnarray*}
GM_{\oplus } &=&3.986004418(9)\times 10^{14}\text{ m}^{3}\text{/s}^{2} \\
R &=& 6.373044737(1)\times 10^{6}
\text{ m} \\
k &=&0.3307(5)
\end{eqnarray*}

$R$ is assumed to be the mean squared radius of an ellipsoid.

The values of the parameters are then

\begin{eqnarray*}
a &=&1.\,\allowbreak 391\,808\,224\,\allowbreak 5(20)\times 10^{-9}
\\
b &=&\allowbreak 2.\,\allowbreak 301\,3\,\allowbreak 26(700)\times 10^{-10}
\end{eqnarray*}

These results do not include the systematic effect due to the non-sphericity of the earth.

\end{document}